\documentclass[a4paper,fleqn,usenatbib]{mnras}

\usepackage{graphicx}
\usepackage{latexsym}
\usepackage{epsfig}
\usepackage{color}
\usepackage{multirow}
\usepackage{float}
\usepackage{url}
\usepackage{amsmath,amssymb,bm,color}
\usepackage{ulem}
\usepackage{amsmath}
\usepackage{gensymb}
\usepackage{booktabs}

\newcommand{\de}{\mathrm d}
\newcommand{\odm}{{\Omega_\mathrm{DM}}}

\newcommand{\mdm}{{m_{\rm DM}}}

\newcommand{\sv}{{\langle\sigma_a v\rangle}}

\newcommand{\g}{$\gamma$}
\newcommand{\Fermi}{{\it Fermi}-LAT }
\newcommand{\nn}{\nonumber}

\def\bea{\begin{eqnarray}}
\def\eea{\end{eqnarray}}
\def\be{\begin{equation}}

\def\ee{\end{equation}}

\title{Bounds on WIMP dark matter from galaxy clusters  at low redshift}

\author[Xiuhui Tan and Manuel Colavincenzo]{Xiuhui Tan$^{3,4,1,2,5}$\thanks{E-mail: tanxh@ihep.ac.cn} and  Manuel Colavincenzo$^{1,2}$\thanks{E-mail: colavincenzo.manuel@gmail.com}
  \\ \\
$^{1}$ Dipartimento di Fisica, Universit\`{a} di Torino, Via Pietro Giuria 1, I-10125, Torino, Italy\\
$^{2}$ INFN -- Istituto Nazionale di Fisica Nucleare, Sezione di Torino, via P. Giuria 1, I-10125 Torino, Italy\\
$^{3}$ Institute of High Energy Physics, Chinese Academy of Sciences, Beijing 100049, China\\
$^{4}$ School of Physical Sciences, University of Chinese Academy of Sciences, Beijing 100049, China\\
$^{5}$ Department of Astronomy, Beijing Normal University, Beijing 100875, China\\
}

\begin{document}
\maketitle

\begin{abstract}
The study of the cross-correlation angular power spectrum between gravitational tracers and electromagnetic signals can be a powerful tool to constrain Dark Matter (DM) microscopic properties. 
In this work we correlate \Fermi\ diffuse \g-ray maps with catalogues of galaxy clusters. To emphasize the sensitivity to a DM signal, we select clusters at low-redshift $0<z<0.2$ and with large-halo mass $M_{500}>10^{13}M_\odot$. The analysis is performed with four catalogues in different wavebands, including infrared, optical and X-rays.
No evidence for a DM signal is identified. 
On the other hand, we derive competitive bounds: the thermal cross-section is excluded at 95\% C.L. for DM masses below 20 GeV and annihilation in the $\tau^+-\tau^-$ channel.

\end{abstract}


\section{Introduction}
\label{sec:intro}

Dark Matter (DM) is one of the main ingredients of the Standard Cosmological Model accounting for more than a quarter of the energy in the Universe. Nevertheless, the nature and physical properties of DM are still mysterious to us. Plenty of DM candidates have been proposed (see. e.g., \cite{2010ARA&A..48..495F}) and, in the case of Weakly Interacting Massive Particles (WIMPs), they have non-negligible interaction with ordinary matter. It is expected that DM could annihilate or decay into standard model particles that will emit \g-ray photons by prompt radiation and inverse Compton scattering. 
From a cosmological point of view, DM is necessary to explain the structure of Universe that we observe today. The distribution and clustering properties of the Large Scale Structure (LSS) are the results of the collapse of baryonic matter in presence of DM. This kind of process, that is called "bottom-up" structure formation, is in agreement with the presence of astrophysical objects, e.g. galaxies and cluster of galaxies, embedded into DM halo. 

As shown in \cite{Camera:2012cj}, one possible way to indirectly determine the properties of the DM particles is to consider the cross-correlation between the so-called unresolved \g-ray background (UGRB) and the weak lensing effect of cosmic shear. We can extend the cross-correlation to a generic unbiased tracer of the large scale matter distribution in the Universe and have a comparable signature. 

The UGRB is the extragalactic \g-ray signal that remains after the removal of the Galactic foreground, coming from the interaction of the cosmic rays with the Galactic interstellar medium and radiation, and the contribution of the resolved \g-ray sources, both point-like and extended. The most accurate measurements of the UGRB comes from the Large Area Telescope (LAT) instrument of the {\it Fermi} satellite~\citep{2015IGRB}.

It is known that the UGRB can be composed by many different contributions. Expected astrophysical sources include blazars (BLZs), e.g. \cite{InoueEtAl2009,AbdoEtAl2010}, star-forming galaxies (SFGs), e.g. \cite{Fermi:2012eba} and misaligned Active Galactic Nuclei (mAGNs), e.g. \cite{DiMauro:2013xta}. Apart from these astrophysical components, as already mentioned, \g-rays can be produced by DM annihilation or decay, in particular in the Galactic and extragalactic (sub)halos. 

As stressed in \cite{Camera:2014rja}, but also in \cite{Fornasa:2015qua} the cross-correlation of the UGRB with other observables can improve and complement the information on the components of the UGRB obtained from the auto-correlation. This kind of analysis can put in evidence the \g-ray component due to DM interaction that would be sub-dominant in the auto-correlation analysis. \\\\
\noindent
In this paper we consider, as tracers of the DM distribution, the largest virialized objects formed by the gravitational instability, the cluster of galaxies. 

Different analysis have already used the cross-correlation of the UGRB with gravitational tracers as indirect probe of DM. Refs. \citep{2014PhRvD..90b3514A,Ando:2014aoa,Regis:2015zka,Cuoco:2015,Shirasaki2015} analyzed catalogues of galaxies, in particular with the aim to increase the sensitivity on DM annihilation signal by dividing the galaxy samples into redshift slices.
Refs. \citep{Camera:2012cj,Fornengo:2014cya,Camera:2014rja,Shirasaki:2014noa,Shirasaki:2016kol,TrosterEtal2017,Shirasaki:2018dkz} focused on the cross-correlation of the UGRB with gravitational lensing, since the latter can be a ``cleaner'' tracer of the DM distribution, and in turn of the emission from DM annihilation.

In this work, we the DM annihilation cross section by measuring the cross angular power spectrum (CAPS) between different galaxy cluster catalogues in different bands (optical, infrared and X-rays) and \g-rays from \Fermi. 
We focus on low redshift since it is where the DM signal is predicted to peak (see, e.g., \cite{Fornengo:2013rga}).
An important step forward of this analysis, compared to the others we have cited, is that the constraints are obtained using an accurate estimation of the CAPS covariance matrix performed using mock realisations of both \g-ray \Fermi\ maps and cluster catalogues. The description on how the mocks are obtained and how the covariance matrix is computed is in the companion paper \cite{companion}.\\\\
\noindent
The paper is organized as following: in section \ref{sec:data} we briefly describe the data used in the analysis; in section \ref{sec:tback} we discuss the basic formalism of the physical models used to compute the theoretical CAPS; the statistical analysis and results are presented in section \ref{sec:analysis}; finally, in section \ref{sec:concl} we draw our conclusions and we discuss the results. 

\section{Data}
\label{sec:data}

In this section we outline the data sets we have used for the cross-correlation analysis: (i) the first 9-years data release of the \g-rays from the \Fermi\ telescope, for which we consider the energy range between 630 MeV and 1 TeV, and (ii) a series of galaxy cluster catalogues in different electromagnetic bands.

\subsection{Fermi-LAT}
\label{sec:fermi}

In this section we just summarize the main properties of the \Fermi \g-ray photon maps we have used for our analysis. For an accurate and extensive description, we refer the reader to \cite{AmmazzalorsoEtAl2018} and \cite{companion}.\\
\noindent
We consider 108 months of \Fermi data (from mission week 9 to week 476). The maps used in the analyses are flux maps obtained by dividing the photon count maps by the exposure maps and the pixel area $\Omega_{\mathrm{pix}}=4\pi/N_{\mathrm{pix}}$. The pixelation is obtained using HEALPix \cite{Healpix2005} with a resolution parameter $N_{\mathrm{side}} = 1024$; this resolution corresponds to 12,582,912 pixels and a mean spacing of $\sim 0.06\degree$. We built the flux maps in 100 energy bins evenly spaced in logarithmic scale between 100 MeV and 1 TeV. As done in \cite{AmmazzalorsoEtAl2018}, then we have joined these 100 micro-bins into 9 larger energy bins, from 631 MeV to 1 TeV, as reported in table \ref{tab:Ebins}.

\begin{table}
    \centering
    \begin{tabular}{cccccc}
        \toprule
        Bin & $\rm E_{\rm min}$ [GeV] & $\rm E_{\rm max}$  [GeV]  & $\ell_{\mathrm{min}}$ & $\ell_{\mathrm{max}}$ & $\theta_{\mathrm{cont}}(\mathrm{deg})$ \\
        \midrule
         1 & 0.631 & 1.202 & 40 & 251  & 0.50  \\ 
         2 & 1.202 & 2.290 & 40 & 316  & 0.58  \\ 
         3 & 2.290 & 4.786 & 40 & 501  & 0.36  \\
         4 & 4.786 & 9.120 & 40 & 794  & 0.22  \\
         5 & 9.120 & 17.38 & 40 & 1000 & 0.15  \\
         6 & 17.38 & 36.31 & 40 & 1000 & 0.12  \\
         7 & 36.31 & 69.18 & 40 & 1000 & 0.11  \\
         8 & 69.18 & 131.8 & 40 & 1000 & 0.10  \\
         9 & 131.8 & 1000  & 40 & 1000 & 0.10  \\
         \bottomrule
    \end{tabular}
    \caption{Energy bins in GeV used in our analysis. $E_{\rm min}$ and $E_{\rm max}$ denote the lower and upper bound of the bins, while $\ell_{\mathrm{min}}$ and $\ell_{\mathrm{max}}$ show the interval in multipole $\ell$ over which the fit of the angular power spectrum is performed: the lower bound is chosen in order to exclude possible galactic-foreground residual contamination, the upper limit is driven by the \Fermi  PSF, whose 68\% containment angle $\theta_{\rm cont}$ is reported.}
    \label{tab:Ebins}
\end{table}

\subsection{Galaxy cluster catalogues}
\label{sec:galaxy}

The galaxy cluster catalogues we have employed for this analysis are the same used in \cite{companion}. We refer to that paper and to the references in table \ref{tab:CatTab} for further details on each catalogue.
\begin{table}
    \centering
    \begin{tabular}{cc}
        \toprule
        Cluster Catalogue & Reference \\
        \toprule
        \textbf{Infrared}\\
        WHY18 & \cite{WHY182018} \\
        \midrule
        \textbf{Optical} \\
        SDSSDR9  & \cite{SDSSDR9_2018} \\
        \midrule
        \textbf{X-Ray} \\
        MCXCsub & \cite{Reiss_2018} \\
        HIFLUGCS & \cite{HIFLUGCS2002} \\
        \bottomrule
    \end{tabular}
    \caption{Galaxy cluster catalogues adopted in the analysis.}
    \label{tab:CatTab}
\end{table}
We consider four catalogues in three different bands. All the galaxy catalogues have been pre-selected so that we keep only clusters with redshift smaller than 0.2, masses larger than $10^{13}M_\odot$ and richness larger than 5.
This is to make the analysis more robust and to focus on the objects that are more interesting for DM searches. 
\begin{itemize}
    \item WHY18 combines photometric galaxies from 2MASS, the Wide-field Infrared Survey Explorer \cite[WISE,][]{WISE2010} and the SuperCOSMOS Sky Survey \citep{SUPERCosmos2001} for a total number of 47,500 clusters~\cite{WHY182018};
    \item SDSSDR9 is a collections of SDSS clusters selected with a model for the galaxy distribution based on cluster density radial profile, the  galaxy luminosity function and the redshift; the total number of clusters for this catalogue is 49,479~\cite{SDSSDR9_2018}:
    \item MCXCsub is built starting the larger catalogue of MCXC \citep{MCXC2011} by selecting a sub-set of 112 clusters with $M_{500}>10^{13}M\odot$, angular diameter larger than 0.2\degree, latitude larger than 20\degree, selected in a portion of the sky that avoids point-source contamination~\cite{Reiss_2018};
    \item HIFLUGCS contains 63 cluster selected from observations with the ROSAT telescope \citep{ROSAT1999}.
\end{itemize}
To avoid systematic effects due to Galactic contamination or cluster misidentification, we mask each of these catalogues in the same way as described at the end of section 2 in \cite{companion}.

\section{Theoretical Background}
\label{sec:tback}

In this section we briefly  describe the theoretical framework behind the physical models considered in our analysis. The basic principles are inherited from \cite{2015ApJS..221...29C} and \cite{BranchiniEtAl2017}. We estimate the CAPS, using the Limber approximation \cite{1953ApJ...117..134L}, by integrating the three dimensional power spectrum $P_{C \gamma}(k)$ and the window function as follows:
\be
 C_\ell^{(C\gamma)}=\int \frac{d\chi}{\chi^2} \ W_{\gamma}(\chi)\, W_{C}(\chi)\,P_{C\gamma}\left(k=\ell/\chi,\chi\right)\,,
\label{eq:clfin}
\ee
where $\chi(z)$ denotes the radial comoving distance, that, 
in a flat cosmology, is given by $\de\chi=c\,dz/H(z)$,  $H(z)$ is the Hubble parameter, $W_{C}(\chi)$ and $W_{\gamma}(\chi)$ are the window functions that characterize, respectively, the redshift distribution of galaxy clusters and \g-ray emitters. Eq. \ref{eq:clfin} is a general expression for the CAPS that we have written for clusters (C) and \g-rays ($\gamma$).

In the following we are going to describe the different contributions that enter in eq. \ref{eq:clfin} and at the end of this section we show the model we have used to fit the measured CAPS.

\subsection{Window function}
\label{sec:windows}
The window function provides us the weights of the signal contribution of a given class of objects at different redshifts. The window functions of the four \g-ray emitters considered in this work (SFGs, BLZ, mAGNs, and DM) are computed as in \cite{2015ApJS..221...29C}; for annihilating DM, the expression is given by:
\begin{eqnarray}
W_{\delta^2}(\chi) & &= \frac{(\odm \rho_c)^2}{4\pi} 
\frac{\sv}{2m_{\mathrm{DM}}^2} \left[1+z(\chi)\right]^3 
\Delta^2(\chi) \nn \\
& &\times \int_{E_{\gamma}>E_{\mathrm{min}}}　\de E_\gamma \, \frac{\de N_{\delta^2}}{\de E_\gamma} 
\left[E_\gamma(\chi) \right] e^{-\tau\left[\chi,E_\gamma(\chi)\right]}\,
\label{eqn:windowanniDM}
\end{eqnarray}
where $\rho_c$ is critical density of the Universe, 
$\odm$ is the cosmological abundance of DM, $\mdm$ and $\sv$ are the DM mass and velocity-averaged annihilation rate,
$\frac{\de N_{\delta^2}}{\de E_\gamma}$ is the energy spectrum of \g-rays originated from DM annihilation and $\tau\left[\chi,E_\gamma(\chi)\right]$ is the \g-ray attenuation function, caused by absorption due to pair-production with extragalactic background light as described in \cite{DomínguezEtAl2012}.

In this work we consider four DM annihilation channels: $\tau^+ \tau^-$, $b\bar b$, $W^+ W^-$ and $\mu^+\mu^-$. The recipe to compute the \g-ray spectrum from these channels is taken from \cite{CirelliEtAl2011}; 

$\Delta^2(\chi)$ denotes the so-called clumping factor:
\begin{equation}
    \begin{split}
       & \Delta^2(z) \equiv 
\frac{\langle \rho^2_{\rm DM} \rangle}{{\bar \rho}^2_{\rm DM}}=\int_{M_{\rm min}}^{M_{\rm max}} \de M \frac{\de n}{\de M}(M,z) \\
&\times \left[1+b_{\rm sub}(M,z)\right]\int \de^3 \mathbf{x} \, 
\frac{\rho^2_h({\mathbf{x}|M,z)}}{{\bar \rho}^2_{\rm DM}}\,.\\
\label{eqn:clumping}
    \end{split}
\end{equation}
This factor describes how the emission is ``boosted'' by the DM clumpiness with respect to the case of a homogeneous density. The halo number density $\frac{\de n}{\de M}(M,z)$ is described as in \cite{Sheth:1999mn} and the mass integral runs from from minimal halo mass $M_{\rm min}=10^{-6}M_{\odot}$ to maximum 
$M_{\rm max}=10^{18}M_{\odot}$. We adopt a NFW halo profile $\rho_h$ from \cite{1996ApJ...462..563N}.
The model for substructures contribution $b_{\rm sub}$ is taken from \cite{Moline2016}.

Note that the window function in eq.~\ref{eqn:windowanniDM} depends on DM mass and annihilation rate: we will derive constraints in the plane $\mdm-\sv$ (taking fixed the other parameters entering in the description of the DM contribution).

For astrophysical \g-ray sources, the window function is characterized by the \g-ray luminosity function $\Phi_{\rm S}$ through the equation: 
\begin{equation}
\begin{split}
 W_{S}(\chi) = &\chi^2(z) \int_{\mathcal{L}_{\rm min}}^{\mathcal{L}_{\rm max}} \de \mathcal{L}
   \, \Phi_{\rm S}(\mathcal{L},z) \\
   \times & \frac{\de N_{\rm S}}{\de E}\left(\mathcal{L},z\right)\times e^{-\tau\left[E(1+z),z\right]}  \ ,
\label{eq:Wastro}
\end{split}
\end{equation}
where $\mathcal{L}$ is the \g-ray rest-frame luminosity (we use the one in the energy interval $0.1$ to $100\,\mathrm{GeV}$), $S$ refers to a generic astrophysical source population and $\de N_{\rm S}/\de E$ is the correspondent energy spectrum. For the description of SFG and mAGN, we follow \cite{Cuoco:2015} while in the case of BLZ, the energy spectrum and the luminosity function are taken from \cite{ajello2015IGRB}.

Finally, for the galaxy clusters, the window function is given by:
\be
W_{C_j}(\chi)=\frac{H(z)}{c}\frac{dN_{C_j}}{dz}\ ,
\label{eqn:Wg}
\ee
with the redshift distribution $dN_{C_j}/dz$ of clusters in the catalogue $j$  taken from \cite{companion}. 
It is worth to briefly describe the nature of the astrophysical terms. For a more extensive review we refer to \cite{Fornasa:2015qua}.

\subsection{3D power spectrum}
\label{sec:poerspectrum}
The power spectrum is the estimate of the amplitude of the fluctuation at a given redshift in Fourier space. Using the halo model approach, the power spectrum can be separated into two contributions: the one-halo and the two-halo terms. In the case of cross-correlation between cluster catalogues and \g-rays from annihilating DM, the two terms can be written as:

 \be
 \begin{split}
 P_{C_j,\delta^2}^{1h}(k,z) &=\int_{M_{\rm min}}^{M_{\rm max}} dM\ \frac{dn}{dM} \frac{\langle N_{C_j}\rangle}{\bar n_{C_j}} \,\frac{\tilde u(k|M)}{\Delta^2} \label{eq:PSannLSS1}
 \end{split}
 \ee
 \be
 \begin{split}
 P_{C_j,\delta^2}^{2h}(k,z) &= \left[\int_{M_{\rm min}}^{M_{\rm max}} dM\,\frac{dn}{dM}b_h(M) \frac{\langle N_{C_j}\rangle}{\bar n_{C_j}}\right] \times \\
 &\left[\int_{M_{\rm min}}^{M_{\rm max}} dM\,\frac{dn}{dM} b_h(M) \frac{\tilde u(k|M)}{\Delta^2} \right]\,P^{\rm lin}(k)\,;
 \label{eq:PSannLSS2}
 \end{split}
 \ee

 The function $\tilde u(k|M)$ is the Fourier transform of the DM emission spatial profile~\cite{Cuoco:2015}, $b_h$ is the halo bias, as defined in  \cite{Sheth:1999mn}, and $P^{\rm lin}(k)$ is the linear power spectrum;
 $\langle N_{C_j}\rangle=(dn_{C_i}/dM)/(dn/dM)$ is the effective halo occupation of clusters~\citep{companion}, that depends on specific catalogue $j$, and the average number density of clusters, at a given redshift z, is computed as $\bar n_{C_j}(z)=\int dM \langle N_{C_j}\rangle\, dn/dM$.

The 3D power spectrum of th cross-correlation between clusters and astrophysical \g-ray sources is given by:
\be
\begin{split}
 P_{C_j,S_i}^{1h}(k,z) &= \int_{\mathcal{L}_{\rm min,i}(z)}^{\mathcal{L}_{\rm max,i}(z)}  d\mathcal{L}\,\Phi_i(\mathcal{L},z)\,\frac{\mathcal{L}}{\langle f_{S_i} \rangle} \,
  \frac{\langle N_{C_j}\!(\mathcal{L})\,\rangle}{\bar n_{C_j}} \label{eq:PSBd1} 
  \end{split}
 \ee
 \be
 \begin{split}
  P_{C_j,S_i}^{2h}(k,z) &=\left[\int_{\mathcal{L}_{\rm min,i}(z)}^{\mathcal{L}_{\rm max,i}(z)} d\mathcal{L}\,\Phi_i(\mathcal{L},z)\, b_{S_i}(\mathcal{L})\,\frac{\mathcal{L}}{\langle f_{S_i} \rangle} \right]\times\\
  &\left[\int_{M_{\rm min}}^{M_{\rm max}} dM\,\frac{dn}{dM} b_h(M)\,\frac{\langle N_{C_j}\,\rangle}{\bar n_{C_j}}  \right] \,P^{\rm lin}(k)\ ,
  \label{eq:PSBd2}
  \end{split}
 \ee

where $\langle f_{S} \rangle=\int d\mathcal{L}\mathcal{L}\Phi_i$, $b_{S}$ is the bias of \g-ray astrophysical sources with respect to matter, for which we adopt $b_{S}(\mathcal{L})=b_h(M(\mathcal{L}))$, and we use $N_{C_j}\!(\mathcal{L})=N_{C_j}(M(\mathcal{L}))$.
The two previous relations require the specification of the relation between the halo-mass and the \g-ray luminosity: we will use the modeling of $M(\mathcal{L})$ derived in \cite{Camera:2014rja}. This relation is rather uncertain for \g-ray objects. While it has a minor impact on the two-halo term (slightly modifying the amplitude of bias), it can dramatically change the size of the one-halo term. On the other hand, the latter is relatively easy to model in an effective way since it does not depend on $k$. To account for this uncertainty we will consider an additional shot-noise term in our model.

\subsection{Full Model}
\label{sec:model}

Using the \g-ray maps described in section \ref{sec:fermi} and the galaxy cluster catalogues described in section \ref{sec:galaxy}, we have measured the CAPS in each of the 9 \g-ray energy bin and for each of the galaxy cluster catalogue. In \cite{companion} we show that this measured signal is compatible with a non-null detection. The model we use to describe the signal takes into account the contributions from the main astrophysical components (BLZs, SFGs, mAGNs) of the UGRB and from DM annihilation. 

We can write our CAPS model in the energy bin $a$ in the following way:
\begin{equation}
\begin{split}
\centering
    &C_{\ell,\mathrm{model}}^{(C\gamma_a)}=\ N_{\mathrm{SN}} \ E^{-\alpha_\mathrm{SN}}_a \ \Delta E_a\\
    &+N_{\mathrm{BLZ}} \ C_{\ell,\mathrm{BLZ}}^{\Delta E_a}+N_{\mathrm{mAGN}} \ C_{\ell,\mathrm{mAGN}}^{\Delta E_a}\\
    &+N_{\mathrm{SFG}} \ C_{\ell,\mathrm{SFG}}^{\Delta E_a}+N_{\mathrm{DM}} \ C_{\ell,\mathrm{DM}}^{\Delta E_a}(M_{\mathrm{DM}})\,,
    \label{eqn:cth}
\end{split}
\end{equation}
with $\Delta E_a$ being the width of the bin and we will call $E_a$ its geometric mean.
The first term is a "shot-noise" flat term modulated in energy by $E_a^{-\alpha_{\mathrm{SN}}}\Delta E_a$. As mentioned above, it accounts for possible uncertainties in the modeling of the $M(\mathcal{L})$ relation for astrophysical \g-ray emitters.

The DM contribution in eq. \ref{eq:clfin} has to be read as the contribution in the case of a specific annihilation channel ($\tau^+ \tau^-$, $b\bar b$, $W^+ W^-$ and $\mu^+\mu^-$). 
We model the DM annihilation term with $N_{\mathrm{DM}} \ C_{\ell,\mathrm{DM}}^{\Delta E_a}(M_{\mathrm{DM}})$, where $N_{\mathrm{DM}}$ is defined as the ratio $\sv/\sv_{0}$ with $\sv_0=3 \times 10^{-26}\mathrm{cm^{3}s^{-1}}$ defining the so-called "thermal" cross-section. 

\begin{figure}
    \centering
    \includegraphics[scale=0.27]{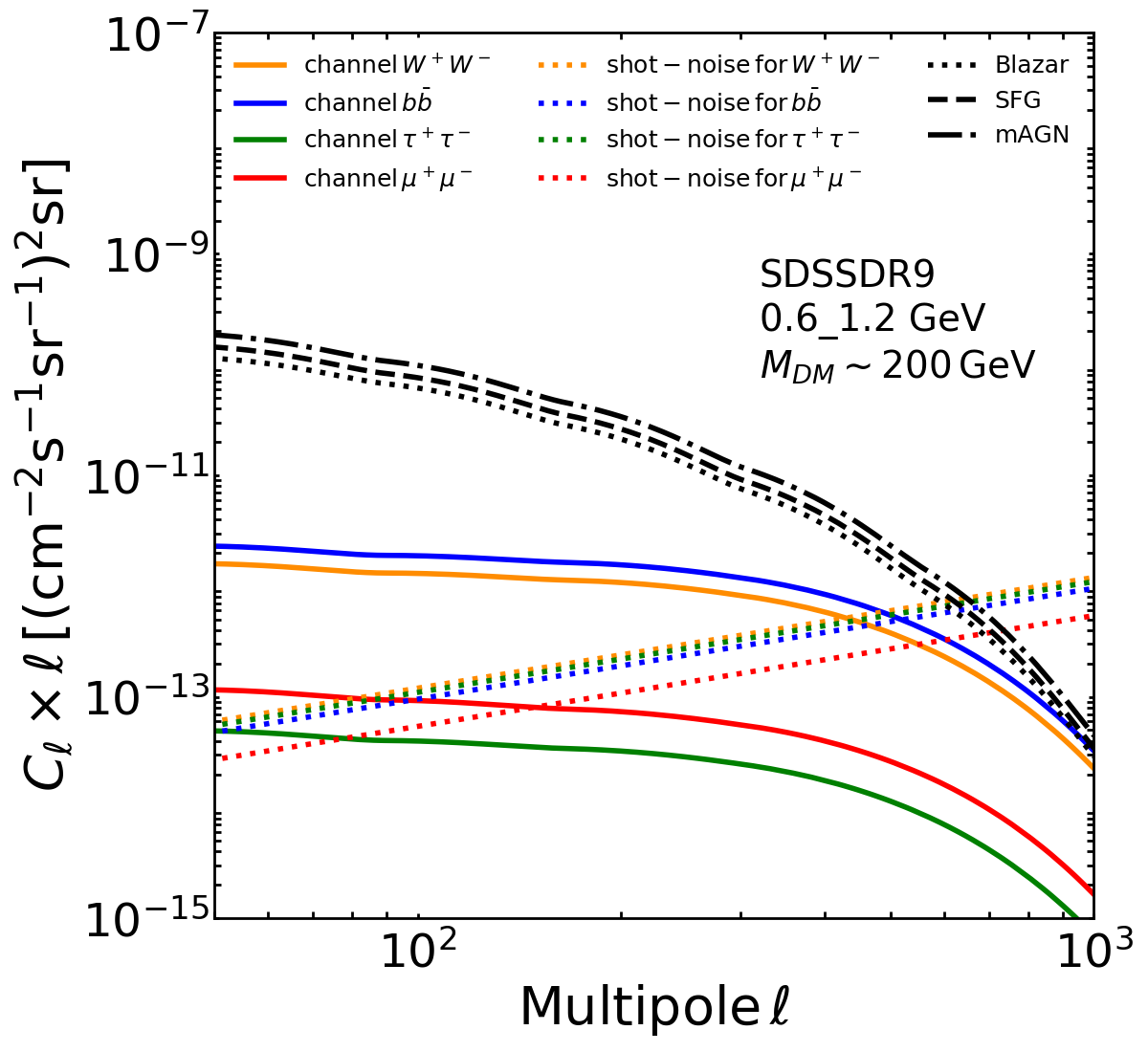}
    \caption{Different contributions to the model of eq. \ref{eqn:cth} in the case of the SDSSDR9 catalogue and the energy bin 0.6-1.2 GeV. For the DM model we considered a mass of $\sim 200$ GeV. The solid lines shows the DM terms,with different colours representing the different annihilation channels. The coloured dotted lines show the shot-noise terms, while in black we report the astrophysical contributions.}
    \label{fig:physicalcl}
\end{figure}

In Figure \ref{fig:physicalcl}, we show all the contributions to the CAPS model, having set the normalization parameters to unity. 
As an example, we consider the case of SDSSDR9 in the 0.6-1.2 GeV energy bin and a DM mass of $ 200$ GeV. The solid lines show the DM terms (with different colours indicating different annihilation channels), the dotted coloured lines show the shot-noise contributions and the black lines show the astrophysical terms. 
As expected the shot-noise is important at very small scale and the astrophysical contributions dominates over the reference DM model at large scales.

\section{Statistical analysis and Results}
\label{sec:analysis}

The comparison of the measured CAPS between the \g-ray maps and cluster catalogues is carried on by defining the $\chi^2$ distribution function:

\begin{equation}
    \chi^2=\sum_{a=1}^{N_{E_\mathrm{bin}}}\sum_{\ell,\ell'}(C_{\ell}^{C\gamma_a}-C_{\ell,\mathrm{model}}^{C\gamma_a})\ [ \Gamma^{C\gamma_a}_{\ell \ell'}]^{-1}
    \ (C_{\ell'}^{C\gamma_a}-C_{\ell',\mathrm{model}}^{C\gamma_a})\,,
    \label{eq:chi2}
\end{equation}

where $N_{E_\mathrm{bin}}$ is the number of energy bin, $\ell$ and $\ell'$ go from $\ell_{\mathrm{min}}$ and $\ell_{\mathrm{max}}$ as indicated in table \ref{tab:Ebins}, $C_{\ell}^{C\gamma_a}$ is the measured CAPS in the ith energy bin and $C_{\ell,\mathrm{model}}^{C\gamma_a}$ is the CAPS model from eq. \ref{eqn:cth}. $[\Gamma^{C\gamma_a}_{\ell \ell'}]^{-1}$ is the inverse of the CAPS covariance matrix within the ith energy bin.

For the estimation of the full CAPS covariance matrix using mock realisations of both \g-maps and cluster catalogues we refer to \cite{companion}. Here we just mention that our covariance matrix neglects the possible covariance between different energy bins; in other words we are considering a non-diagonal covariance matrix in angular scale and a diagonal matrix in energy. As shown in \cite{companion}, we can neglect the non-diagonal correlation of the errors between different energy bins because they are smaller than 5\% for every angular scale.
\begin{figure*} 
    \centering
    \includegraphics[scale=0.42]{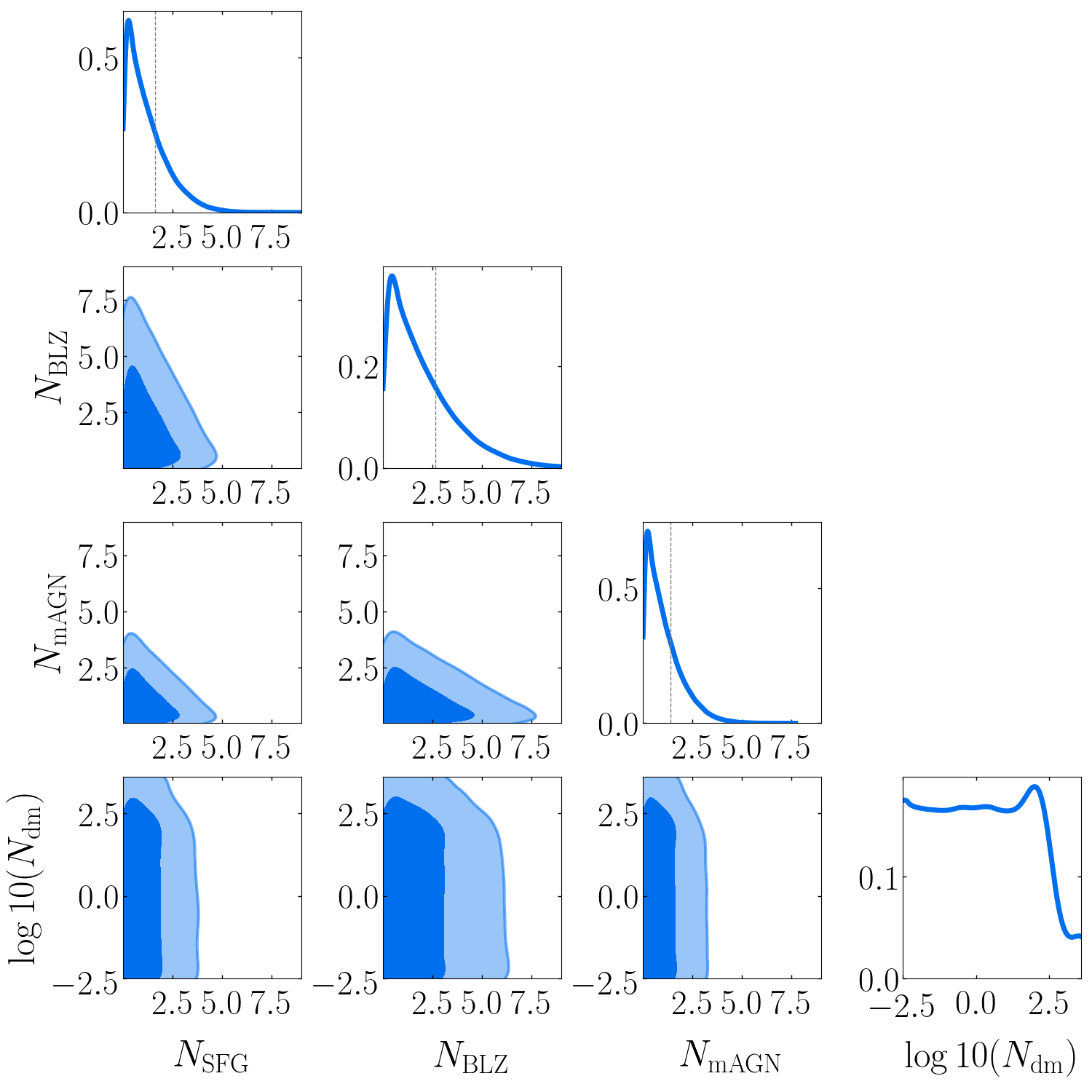}
    \caption{Triangle plot reporting the posterior distribution for the astrophysical parameters and DM cross-section in the case of the $W^+W^-$ annihilation channel and for the WHY18 cluster catalogue. The dark blue region and light blue region denote the $95\%$ and $68\%$ C.L., respectively.}
    \label{fig:old3suv_ww}
\end{figure*}

The computation of the $\chi^2$, shown in eq. \ref{eq:chi2}, depends on the 7D parameter vector given by 
$P=(N_{\mathrm{SN}},\, \alpha_{\mathrm{SN}},\, N_{\mathrm BLZ},\, N_{\mathrm{mAGN}},\, N_{\mathrm{SFG}}, \,N_{\mathrm{DM}}, \,M_{\mathrm{DM}})$. 
The normalization of the astrophysical ($N_{\mathrm BLZ},\, N_{\mathrm{mAGN}}$ $N_{\mathrm{SFG}}$) and the shot-noise ($N_{\mathrm{SN}}$) contributions are sampled in linear scale, as well as the spectral index of the shot-noise ($\alpha_{\mathrm{SN}}$), while the DM parameters ($N_{\mathrm{DM}}$, and $M_{\mathrm{DM}}$) are sampled in logarithm scale. 
We adopted a Markov Chain Monte Carlo (MCMC) routines provided by the MontePython package\footnote{https://monte-python.readthedocs.io/en/latest/} to scan the multi-dimensional parameter space. 
The convergence of the chains are controlled bu the Gelman-Rubin criterion.

The triangle plot in figure \ref{fig:old3suv_ww} shows an example of the outcome of our statistical analysis. We report the probability distributions of the normalization parameters for DM and astrophysical sources in the case of the catalogue WHY18 and for a DM model annihilating into $W^{+} W^{-}$ final state.

Dark blue regions identify the 95\% C.L. and light blue is for 68\% C.L.. 
The dotted vertical lines in the marginal 1D distributions show the 68\% C.L. limits.
From this figure it appears to be clear that we can only put upper bounds on the astrophysical components and on the DM cross-section. This remains true for all the annihilation channels and for the other cluster catalogues.

In table \ref{tab:fitvalue} we report the most contraining 68\% C.L. upper limits for the normalization parameters of the astrophysical components among the various cases considered in the analysis.
\begin{table} 
    \centering
    \begin{tabular}{|c|c|c|}
        \toprule
        \multicolumn{2}{|c|}{68\% C.L. upper bounds}\\
        \midrule
        WHY18 & $N_{\mathrm{SFG}}$ & 1.6  \\
        WHY18 & $N_{\mathrm{BLZ}}$ & 2.7  \\
        HIFLUGCS & $N_{\mathrm{mAGN}}$ & 1.0  \\
        \bottomrule
     \end{tabular}
        \caption{Best upper limits on the astrophysical parameters. The catalogue from which the bound has been derived is indicated within brackets.}
        \label{tab:fitvalue}
\end{table}
Table \ref{tab:chi2Tab} shows the $\chi^2$ values of the best-fit models for the various catalogues and annihilation channels.  

\begin{figure*}
    \centering
    \includegraphics[scale=0.53]{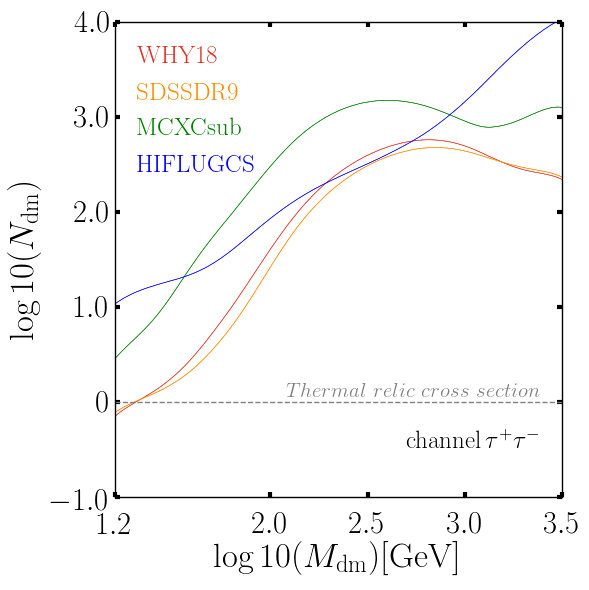}
    \includegraphics[scale=0.53]{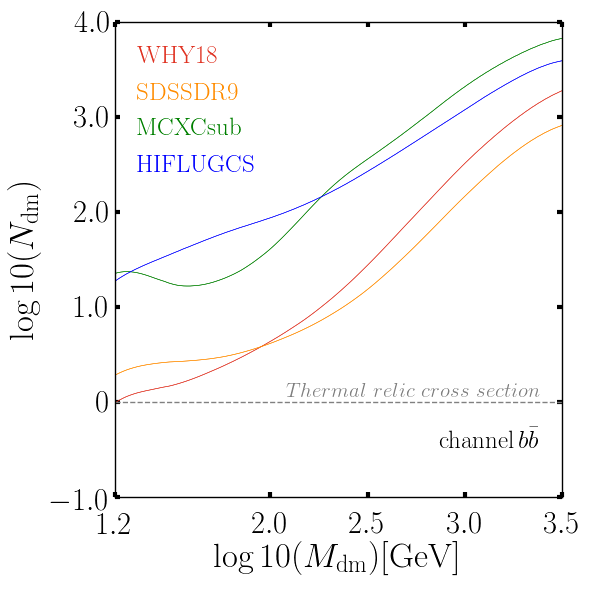}
    \includegraphics[scale=0.53]{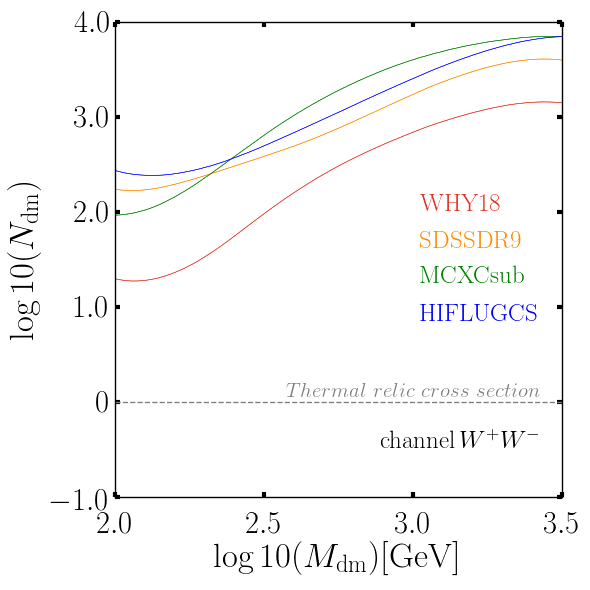}
    \includegraphics[scale=0.53]{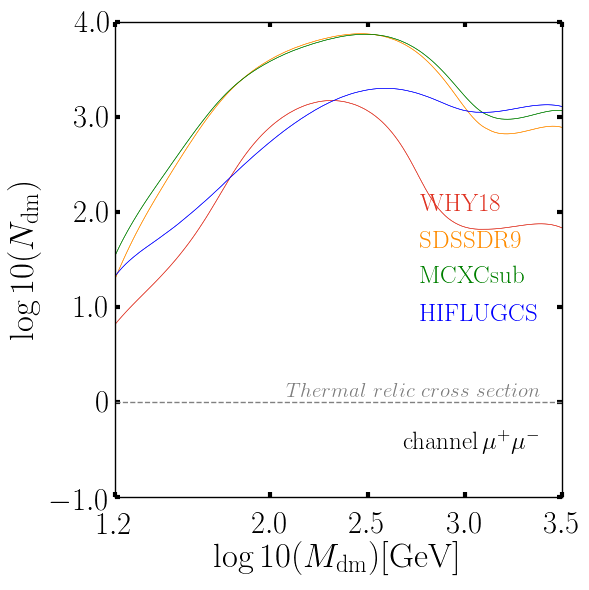}    
    \caption{$95\%$ C.L. upper bounds on the DM annihilation rate ratio $N_{DM}=\langle \sigma v\rangle/\langle \sigma v\rangle_0$ as a function of the DM Mass for the four annihilation final state under consideration (upper left $\tau^+\tau^-$, upper right $b\bar b$, bottom left $W^+ W^-$ and bottom right $\mu^+ \mu^-$).}
    \label{fig:channeldm}
\end{figure*}

In Figure \ref{fig:channeldm} we show the 95\% C.L. upper limits on the DM cross-section as function of the DM mass for the four annihilation channels.
\begin{table}
    \centering
    \begin{tabular}{|c|c|c|c|c|}
    \toprule
       &  $\chi^2$($\tau^+ \tau^-$)  & $\chi^2$($b\bar b$) & $\chi^2$($W^+ W^-$) & $\chi^2$($\mu^+\mu^-$)\\
    \midrule
         WHY18&  91.01&91.76&91.03&91.39\\
         SDSSDR9&109.5&110.5&109.5&109.7\\
         MCXCsub&137&137.1&139.2&137.8\\
         HIFLUGCS&91.46&91.16&91.58&90.84\\    
    \bottomrule            
    \end{tabular}
    \caption{$\chi^2$ of the best-fit model for all catalogues and annihilation channels considered in this analysis.}
    \label{tab:chi2Tab}
\end{table}

\begin{figure}
    \centering
    \includegraphics[scale=0.5]{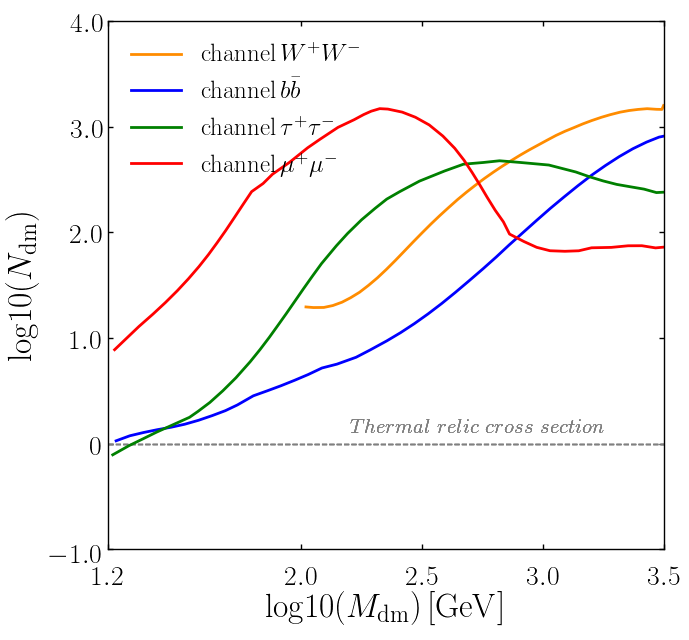}
    \caption{Most constraining bound, at each DM mass, among the four cases in Figure~\ref{fig:channeldm}. We show $W^+ W^-$(yellow), $b\bar b$(blue), $\tau^+\tau^-$(green), and $\mu^+ \mu^-$(red) annihilation channels.}
    \label{fig:clusterdm}
\end{figure}

\section{Discussion and Conclusions}
\label{sec:concl}

In this work, we have analyzed the cross angular power spectrum between the UGRB observed by
\Fermi  and four galaxy clusters catalogues to constrain the properties of WIMP DM.

Here below the main conclusions that can be drawn from the results shown in the previous section.

\begin{itemize}
\item Astrophysical sources \\
Due to the small volume ($z<0.2$) and the relative low number of objects in the catalogues, considered in our analysis, only upper bounds can be derived for the contributions from \g-ray astrophysical sources.
Blazars are only weakly constrained ($N_{\mathrm{BLZ}}\le 2.7$ at 68\% C.L.) since their unresolved component is predominantly located at higher redshift.
More stringent bounds can be obtained for SFG ($N_{\mathrm{SFG}}\le 1.6$ at 68\% C.L.) and, in particular, for mAGN ($N_{\mathrm{mAGN}}\le 1.0$ at 68\% C.L.), which are believed to be the main contributor of the UGRB in the Local Universe~\citep{AmmazzalorsoEtAl2018}.

\item DM - comparison among different catalogues \\
From figure \ref{fig:channeldm}, one can note that the most constraining catalogues are WHY18 and SDSSDR9. This can be understood by considering that with respect to X-ray catalogues, they have much more objects (and thus statistics), whilst, on the other hand, a weaker detected signal~\citep{companion}.

\item DM - comparison among different channels \\
In Figure \ref{fig:clusterdm}, we summarize our finding for DM, by reporting the combined bounds for each annihilation channel. The curves are derived from Figure \ref{fig:channeldm} by taking the most constraining catalogue at each DM mass. 
At low DM masses, the dominant mechanisms of \g-ray production is through prompt emission occurring mostly via $\pi^0$-decay for the cases of $b \bar b$ 
$\tau^+ \tau^-$ and $W^+ W^-$, while through final state radiation for $\mu^+ \mu^-$.
The latter case is the one that is less constrained below 100 GeV.

At high energies, the inverse Compton of CMB photon with electrons and positrons generated by DM annihilation can provide a sizeable \g-ray contribution.
The injection of electrons and positrons is enhanced in leptonic channels with respect to $b \bar b$ and $W^+ W^-$ channels and, for this reason, $\mu^+ \mu^-$ is the most constraining final state at TeV energies (followed by the other leptonic channel, $\tau^+ \tau^-$).

\end{itemize}

Summarizing, through the analysis of clusters at low-redshift $0<z<0.2$, with large-halo mass $M_{500}>10^{13}M_\odot$, and in three different wavebands (infrared, optical and X-rays), we excluded the thermal cross-section of DM at 95\% C.L. for DM masses below 20 GeV (15 GeV) and annihilation in the $\tau^+-\tau^-$ ($b \bar b$) channel. 

The current work has been conducted treating all the clusters in the same way, i.e., without introducing any weight for the clusters.
We plan to increase the sensitivity to DM in a future improvement of the work, by statistically weighting the clusters according to the expected annihilation signal.

\section*{Acknowledgements}

This work is supported by the following grants:  {\sl Departments of Excellence} (L. 232/2016), awarded by the Italian Ministry of Education, University and Research (MIUR); {\sl The Anisotropic Dark Universe}, Number CSTO161409, funded by Compagnia di Sanpaolo and University of Torino; {\sl TAsP (Theoretical Astroparticle Physics)} project, funded by the Istituto Nazionale di Fisica Nucleare (INFN); PRIN 2017 project (Progetti di ricerca di Rilevante Interesse Nazionale) {\sl The Dark Universe: A Synergic Multimessenger Approach}, Number 2017X7X85K, funded by MIUR; ``Deciphering the high-energy sky via cross correlation'' funded by Accordo Attuativo ASI-INAF n. 2017-14-H.0.

\bibliographystyle{mnras}
\bibliography{x-correlation}
\end{document}